\documentclass[aip,jcp,citeautoscript,nofootinbib,reprint,groupedaddress]{revtex4-1}
\usepackage[symbol]{footmisc}
\usepackage[utf8]{inputenc}
\usepackage{color}
\usepackage{ulem}
\usepackage{graphicx}
\usepackage{dcolumn}
\usepackage{bm}
\usepackage{epsfig}
\usepackage{epstopdf}
\usepackage{amsmath}
\usepackage{amssymb}
\usepackage{amsfonts}  
\usepackage{bm}
\usepackage{braket,tikz}
\usetikzlibrary{shapes,arrows,chains,matrix,positioning,scopes,decorations.pathreplacing,angles,quotes}

\newcommand{\znot}{\mathbf{z}_0^{}}
\newcommand{\znotp}{\mathbf{z}_0^\prime}
\newcommand{\bz}{\mathbf{z}}
\newcommand{\zt}{\mathbf{z}_t^{}}
\newcommand{\ztp}{\mathbf{z}_t^\prime}

\newcommand{\cq}{\mathbf{c}_{\mathbf{q}}}
\newcommand{\cpp}{\mathbf{c}_{\mathbf{p}}}
\newcommand{\pt}{\mathbf{p}_t}
\newcommand{\qt}{\mathbf{q}_t}
\newcommand{\ptp}{\mathbf{p}^\prime_t}
\newcommand{\qtp}{\mathbf{q}^\prime_t}

\newcommand{\mqq}{\mathbf{M}_{qq}}
\newcommand{\mqp}{\mathbf{M}_{qp}}
\newcommand{\mpq}{\mathbf{M}_{pq}}
\newcommand{\mpp}{\mathbf{M}_{pp}}

\newcommand{\mqqbb}{\mathbf{M}_{qq}^b}
\newcommand{\mqpbb}{\mathbf{M}_{qp}^b}
\newcommand{\mpqbb}{\mathbf{M}_{pq}^b}
\newcommand{\mppbb}{\mathbf{M}_{pp}^b}
\newcommand{\mqqf}{\mathbf{M}_{qq}^f}
\newcommand{\mqpf}{\mathbf{M}_{qp}^f}
\newcommand{\mpqf}{\mathbf{M}_{pq}^f}
\newcommand{\mppf}{\mathbf{M}_{pp}^f}
\newcommand{\gamz}{\bm{\gamma}_0}
\newcommand{\gamt}{\bm{\gamma}_t}
\newcommand{\opA}{\hat{A}}
\newcommand{\opB}{\hat{B}}
\newcommand{\opH}{\hat{H}}
\newcommand{\onehalf}{\frac{1}{2}}
\newcommand{\Ronehalf}{^{\frac{1}{2}}}
\newcommand{\RMonehalf}{^{-\frac{1}{2}}}
\newcommand{\bc}{\mathbf{c}}
\newcommand{\T}{^T}
\newcommand{\Kmat}{\mathbf{K}}
\newcommand{\Jmat}{\mathbf{J}}

\newcommand{\Mmat}{\mathbf{M}}
\newcommand{\bx}{\mathbf{x}}
\newcommand{\nullmat}{\mathbb{O}}
\newcommand{\Ximat}{\mathbf{\Xi}}

\newcommand{\Lammat}{\mathbf{\Lambda}}
\newcommand{\Ommat}{\mathbf{\Omega}}

\newcommand{\Amat}{\mathbf{A}}
\newcommand{\Bmat}{\mathbf{B}}
\newcommand{\Cmat}{\mathbf{C}}
\newcommand{\Dmat}{\mathbf{D}}

\newcommand{\Xmat}{\mathbf{X}}
\newcommand{\Ymat}{\mathbf{Y}}

\newcommand{\Imat}{\mathbb{I}}
\newcommand{\Gmat}{\mathbf{G}}

\newcommand{\eqn}[1]{Eq.~\ref{#1}}
\newcommand{\figr}[1]{Fig.~\ref{#1}}

\newcommand{\tabl}[1]{Table~\ref{#1}}

\newcommand{\parenth}[1]{\left(#1\right)}
\newcommand{\brak}[1]{\left[#1\right]}

\begin{document}
\preprint{AIP/123-QED}

\title{Semiclassical dynamics in the mixed quantum-classical limit}

\author{Matthew S. Church}\author{Nandini Ananth}\email{na346@cornell.edu}
\affiliation{Department of Chemistry and Chemical Biology, 
Cornell University, Ithaca, New York, 14853, USA}

\date{\today}
\begin{abstract} 
The semiclassical Double Herman-Kluk Initial Value Representation 
is an accurate approach to computing quantum real time correlation functions,
but its applications are limited by the need to evaluate an oscillatory 
integral. In previous work, we have shown that this `sign problem' 
can be mitigated using the modified Filinov filtration technique to 
control the extent to which individual modes of the system contribute to the 
overall phase of the integrand.
Here we follow this idea to a logical conclusion: 
we analytically derive a general expression for the mixed quantum-classical limit of 
the semiclassical correlation function \textemdash\, AMQC-IVR, 
where the phase contributions from the `classical' modes of the system are
filtered while the `quantum' modes are treated in the full semiclassical limit.
We numerically demonstrate the accuracy and efficiency of the AMQC-IVR formulation 
in calculations of quantum correlation functions and reaction rates using three model 
systems with varied coupling strengths between the classical and quantum 
subsystems.  We also introduce a separable prefactor approximation that further reduces
the computational cost, but is only accurate in the limit of weak coupling between 
the quantum and classical subsystems.

\end{abstract}

\maketitle

\section{\label{sec:level1}Introduction} 
Semiclassical methods based on the initial value representation (SC-IVR) can be used to characterize 
quantum mechanical effects of many-body systems in real time.\cite{mil01a,tho04a,kay05a,her84a,kay94a,kay94b,kay94c} 
Using classical trajectories from molecular dynamics simulations, 
SC-IVR methods accurately describe bound-state motion, tunneling processes, 
chemical reaction rates, and coherence effects in both adiabatic and nonadiabatic 
systems.\cite{ven07a,kay97a,sun98b,zha04a,buc18a,zha04b,moi09a,elr02a,ski99a,sto97a,cor00a,ana07a,mil12b} 
And while there are other classes of trajectory-based methods such as ring polymer molecular 
dynamics\cite{cra04a,cra05a,men11a,hab13a,men14a} and centroid molecular dynamics\cite{cao94a,jan99a} 
that can capture some quantum effects in condensed phase systems, these methods
cannot be used for systems where quantum coherence effects play a role.
Yet other classes of methods are derived from either the exact path integral representation 
of the propagator,~\cite{mak95a} the quantum Liouvillian,~\cite{kap99a} or 
wavepacket dynamics,~\cite{bec00a,mey09a} through a series of rigorous 
approximations.
However, these approaches remain limited 
to low-dimensional systems or condensed phase systems where a large number 
of near-classical modes serve to mitigate the importance of 
long-lived quantum coherence effects.~\cite{kap06a,mak15a,wal15a,wan03a}

Efforts to make SC-IVR methods computationally feasible focus on dealing 
with the `sign problem' that arises from the inclusion of a phase from individual
 trajectories that must be averaged over.
There are a number of existing approximations that make SC-IVR theory more 
amenable to large-scale simulation, such as the widely-used linearized SC-IVR,\cite{wan98a,sun98c,liu15a,shi03a}  
a classical limit of SC theory that is accurate on short time scales but 
suffers from zero-point energy leakage and fails to describe long-time coherence effects.\cite{hab09a,buc18a,mil01a,gel01a,ana07a} 
Other methods include the various forward-backward\cite{sun99a,wan00a,wan01a,gel01a,tho01a} SC-IVRs, and 
there are promising semiclassical methods based on time-averaging,\cite{kal03a,buc18b} 
the ``divide and conquer" methodology,\cite{ceo17a,dil18a,dil18b} linearization,\cite{lee16a} 
symmetrical windowing,\cite{cot13a} and semiclassical quantization\cite{lor17a} for the 
calculation of electronic coherence and/or 1D and 2D vibrational and vibronic spectra. 
Still, however, there is need of practical methods that include a true description of 
nuclear coherence in real time for the study of interesting processes
such as the generation of hot-electrons at metal surfaces, 
intramolecular vibrational relaxation, molecular collisions, and 
other electronically and vibronically nonadiabatic processes.\cite{uze91a,gol15a,kru15a,dom12a,ree09a,gra96a,ham10a}

The mixed quantum-classical IVR (MQC-IVR) is a semiclassical approach to 
calculating correlation functions that has shown promise in mitigating 
the SC-IVR sign problem in low-dimensional adiabatic and nonadiabatic 
systems.\cite{ant15a,chu17a,chu18a} It is derived by applying modified 
Filinov filtration\cite{fil86a,mak87a,bre97a,wan01b,spa05a} (MFF) to
the double Herman-Kluk (DHK-IVR) formulation of the correlation function,
making the level of theory used on each dof tunable (via an adjustable `tuning' parameter) 
between classical and quantum limits of SC-IVR theory.\footnote[2]{Throughout the study 
we may refer to dofs treated in the 
classical limit of SC-IVR theory as `classical', and to dofs treated 
in the quantum limit of SC-IVR theory as `quantum' in order to distinguish 
the level of SC-IVR theory used for different modes in a given system. 
We do this with the understanding that the whole system is described
at the semiclassical level of theory.}
The zero-limit of the tuning parameters is equivalent 
to treating the full system at the DHK-IVR level of theory 
(i.e. the quantum limit SC correlation function), 
and setting the tuning parameters to infinity results
in a classical limit description of the system similar
to LSC-IVR. 
MQC-IVR thus provides a uniform framework for SC simulations 
with mode-specific quantization and no uncontrolled approximations 
to the forces between quantum and classical subsystems, 
as are made in standard multi-physics approaches.\cite{car99a}

A challenge with MQC-IVR is, however, determining the optimal set 
of tuning parameters to minimize both computational cost and 
loss of accuracy. One approach to this challenge, 
and the subject of this study, is to \textit{analytically} evaluate the
general MQC-IVR limit where the tuning parameters associated with the quantum subsystem 
go to zero, and the tuning parameters associated with the classical subsystem 
go to infinity. The result is an analytical mixed quantum-classical
(AMQC-IVR) expression for the SC-IVR time correlation function that 
offers reduced computational effort, and circumvents
the need to find optimal values of the tuning parameters used in MQC-IVR.

In this study we use three multidimensional model systems to 
demonstrate that AMQC-IVR accurately describes quantum dynamical
features of systems over a wide range of coupling strengths between 
quantum and classical subsystems. We also introduce a 
separable prefactor (SP) approximation to the AMQC-IVR prefactor that 
further reduces computational cost. We show that the SP approximation 
is increasingly accurate as the coupling between quantum and classical subsystems decreases,
and increasingly efficient when the classical subsystem is larger than the quantum subsystem. 
Finally, we show that AMQC-IVR has the potential to be systematically improved, 
and to be amenable to a variety of existing representations and approximations in the SC 
literature.\cite{gel00a,dil16a} 

This paper is organized as follows. In Sec.~\ref{sec:level2} we briefly review MQC-IVR theory 
and provide an overview of the derivation of AMQC-IVR. In Sec.~\ref{sec:modsys} 
we describe the model systems and in Sec.~\ref{sec:simdet} we provide simulation details. 
Results are discussed in Sec.~\ref{sec:res} and conclusions are drawn in Sec.~\ref{sec:conc}.

\section{\label{sec:level2}Theory}

\subsection{MQC-IVR}
Throughout this manuscript we use atomic units and take $\hbar=1$. The general MQC-IVR correlation function\cite{chu17a,chu18a}
is given by
\begin{align}
C_{AB}(t)=&\frac{1}{\parenth{2\pi}^{2N}}\int d\znot\int d\znotp A_{\znot\znotp}B_{\ztp\zt}\nonumber\\
\times &e^{i\brak{S_t\parenth{\znot}-S_t\parenth{\znotp}}}D_t\parenth{\znot,\znotp;\bc}e^{-\onehalf\mathbf{\Delta}_0\T\bc\mathbf{\Delta}_0^{}},
\label{eq:MQC1}
\end{align}
where $N$ is the dimensionality of the entire system, and we use the double-forward formulation.~\cite{chu17a}
The $2N$-dimensional phase space vectors of the 
unprimed and primed trajectories at time $t$ are defined as
\begin{align}
\zt=&\parenth{\pt,\qt}\nonumber\\
=&\parenth{p_{t_1}^{},\dots,p_{t_N}^{},q_{t_1}^{},\dots,q_{t_N}^{}}\nonumber\\
=&\parenth{z_{t_1}^{},\dots,z_{t_N}^{},z_{t_{N+1}}^{},\dots,z_{t_{2N}}^{}},\label{eq:ztref}\\
\ztp=&\parenth{\ptp,\qtp}\nonumber\\
=&\parenth{p_{t_1}^\prime,\dots,p_{t_N}^\prime,q_{t_1}^\prime,\dots,q_{t_N}^\prime}\nonumber\\
=&\parenth{z_{t_1}^\prime,\dots,z_{t_N}^\prime,z_{t_{N+1}}^\prime,\dots,z_{t_{2N}}^\prime},
\end{align}
respectively.
We also take $S_t\parenth{\bz}$ to be the classical action of a trajectory originating at 
point $\bz$, and $\bm{\Delta}_0^{}=\znotp-\znot$ is the phase space displacement between 
pairs of forward trajectories at $t=0$. The position-space wavefunction of the coherent state $\ket{\zt}$ at time $t$ is given by
\begin{align}
\braket{\bx|\zt}=\parenth{\frac{\det\brak{\gamt}}{\pi^N}}^\frac{1}{4}e^{-\onehalf\parenth{\bx-\qt}\T\gamt\parenth{\bx-\qt}+i\pt\T\parenth{\bx-\qt}},
\end{align}
where $\gamt$ is a diagonal $N\times N$ matrix that determines the width of the wavepacket. We also represent the coherent state matrix element of a given quantum mechanical operator $\hat{\Omega}$ as
\begin{align}
\Omega_{\bz^{}\bz^\prime}=\braket{\bz^{}|\hat{\Omega}|\bz^\prime}.
\end{align}
A detailed form of the MQC-IVR prefactor $D_t\parenth{\znot,\znotp;\bc}$ is provided in Appendix \ref{ap:dfp}. The diagonal $2N\times 2N$ matrix
of tuning parameters $\bc$ is given by
\begin{align}
\bc=\begin{pmatrix}
\cpp	&	\nullmat	\\
\nullmat	&	\cq
\end{pmatrix},
\label{eq:cmatrix1}
\end{align}
and $\nullmat$ is the null matrix. The elements of the diagonal $N\times N$ matrices $\cpp$ and $\cq$ determine the extent of separation (in momentum and position space, respectively) between trajectory pairs at time $t=0$ and, therefore, determine the extent of phase cancellation in the integrand of \eqn{eq:MQC1}. In the 
limit that all the elements of $\bc$ approach zero, the effect of MFF is removed and \eqn{eq:MQC1} reduces to DHK-IVR,
\begin{align}
C_{AB}(t)=&\frac{1}{\parenth{2\pi}^{2N}}\int d\znot \int d\znotp A_{\znot\znotp}B_{\ztp\zt}\nonumber\\
\times& C_t^{}\parenth{\znot}C_t^{*}\parenth{\znotp}e^{i\brak{S_t\parenth{\znot}-S_t\parenth{\znotp}}},
\label{eq:DHK}
\end{align}
which we refer to as the quantum limit SC-IVR correlation function.
In \eqn{eq:DHK}, $C_t^{}\parenth{\znot}$ is the Herman-Kluk prefactor for a trajectory beginning at point $\znot$,
\begin{align}
C_t\parenth{\znot}=\det\bigg[&\onehalf\big(\gamz\Ronehalf\mqq\gamt\RMonehalf+\gamz\RMonehalf\mpp\gamt\Ronehalf\nonumber\\
&-i\gamz\Ronehalf\mqp\gamt\Ronehalf+i\gamz\RMonehalf\mpq\gamt\RMonehalf\big)\bigg]\Ronehalf,
\label{eq:HKpref}
\end{align}
and elements of the monodromy matrix $\Mmat$ are defined by $\Mmat_{\alpha\beta}=\frac{\partial\bm{\alpha}_t}{\partial\bm{\beta}_0}$ with $\parenth{\bm{\alpha},\bm{\beta}}\in\parenth{\mathbf{p},\mathbf{q}}$.
In the limit that all the elements of $\bc$ approach infinity, the unprimed and primed trajectories are constrained to be identical, resulting in complete phase cancelation in the MQC-IVR integrand. In this limit, MQC-IVR is identical to a classical limit SC-IVR correlation function similar to LSC-IVR,
\begin{align}
C_{AB}(t)=\frac{1}{\parenth{2\pi}^N}\int d\znot\,A_{\znot\znot}B_{\zt\zt},
\label{eq:HUSIVR}
\end{align}
which we refer to as Husimi-IVR. It is clear that the values of $\bc$ in MQC-IVR control the phase contributions
to the integrand from each dof and, therefore, the level of SC theory used to describe each dof.

\subsection{AMQC-IVR}
Here we outline the AMQC-IVR derivation and provide details in Appendix \ref{ap:mlp}. First, as a matter of bookkeeping, we consider a general system with $F$ quantum dofs and $N-F$ classical dofs. We also order the elements of the position and momentum vectors ($\qt$, $\qtp$, $\pt$, and $\ptp$) such that the $F$ elements of the quantum subsystem are listed before the $N-F$ elements of the classical subsystem.

The $\bc$-dependence of the MQC-IVR integrand in \eqn{eq:MQC1} can be written as
\begin{align}
G_t\parenth{\znot,\znotp;\bc}=&D_t\parenth{\znot,\znotp;\bc}e^{-\onehalf \mathbf{\Delta}_0\T\bc\mathbf{\Delta}_0^{}}.
\label{eq:ML1}
\end{align}
We now evaluate \eqn{eq:ML1} 
in the limit that the elements of $\bc$ associated with the quantum subsystem go to zero,
and the limit that the elements of $\bc$ associated with the classical subsystem go to infinity. 
In order to distinguish the phase space variables of the two subsystems, we introduce the following $2F$-dimensional vectors to represent the initial conditions of the quantum subsystem,
\begin{align}
{\bz}_Q^{}=&\parenth{p_{0_1}^{},\dots,p_{0_F}^{},q_{0_1}^{},\dots,q_{0_F}^{}},\\
{\bz}_Q^\prime=&\parenth{p_{0_1}^\prime,\dots,p_{0_F}^\prime,q_{0_1}^\prime,\dots,q_{0_F}^\prime},
\end{align}
and the following $2(N-F)$-dimensional vectors to represent the initial conditions of the classical subsystem,
\begin{align}
{\bz}_C^{}=&\parenth{p_{0_{F+1}}^{},\dots,p_{0_N}^{},q_{0_{F+1}}^{},\dots,q_{0_N}^{}},\\
{\bz}_C^\prime=&\parenth{p_{0_{F+1}}^\prime,\dots,p_{0_N}^\prime,q_{0_{F+1}}^\prime,\dots,q_{0_N}^\prime}.
\end{align}
Using the following $\delta$-function identity,
\begin{align}
\delta(x)=\lim_{a\rightarrow\infty}\parenth{\frac{a}{2\pi}}\Ronehalf e^{-\frac{a}{2}x^2},
\label{eq:ML2}
\end{align}
we obtain
\begin{widetext}
\begin{align}
\lim_{c_{\text{quantum}}\rightarrow0\atop c_{\text{classical}}\rightarrow\infty} G_t\parenth{\znot,\znotp;\bc}=&\parenth{2\pi}^{N-F}C_t^{}\parenth{\znot}C_t^*\parenth{\znotp}\Lambda_t\parenth{\znot,\znotp}
\prod\delta\parenth{z_C^\prime-z_C^{}}.
\label{eq:ML3}
\end{align}
\end{widetext}
Note that the product on the right-hand side of \eqn{eq:ML3} is over all $2\times(N-F)$ elements of the classical subsystem in $\bz_C^{}$ and $\bz_C^\prime$. In addition, we note that the prefactor can be written as a product of a pair of Herman-Kluk prefactors
for the primed and unprimed trajectories and $\Lambda_t\parenth{\znot,\znotp}$, an additional term that accounts 
for coupling between the quantum and classical subsystems defined in Appendix \ref{ap:mlp}. 
Substituting the limit of \eqn{eq:ML3} into \eqn{eq:MQC1}, and then evaluating the integrals over the primed initial conditions of the classical subsystem, i.e. over $d\bz_C^\prime$,
we obtain the AMQC-IVR correlation function,
\begin{align}
C_{AB}(t)=&\frac{1}{\parenth{2\pi}^{N+F}}\int d\znot\int d\bz_Q^\prime A_{\znot\bar{\bz}_0^{}}B_{\zt\ztp}\nonumber\\
\times &C_t^{}\parenth{\znot}C_t^*\parenth{\bar{\bz}_0^{}}e^{i\brak{S_t\parenth{\znot}-S_t\parenth{\bar{\bz}_0}}}\nonumber\\
\times & \Lambda_t\parenth{\znot,\bz_Q^\prime}.
\label{eq:ML4}
\end{align}
The $2N$-dimensional phase space vector $\bar{\bz}_0$ replaces the classical components of $\znotp$ with the classical components of $\znot$.

In obtaining \eqn{eq:ML4} from \eqn{eq:MQC1} we have constrained the initial conditions of the classical subsystem in the unprimed and primed trajectories to be identical, and, consequently, reduced the dimensionality of the phase space integral by $2\times(N-F)$, twice the dimensionality of the classical subsystem. The additional prefactor $\Lambda_t$ appears as a result of this constraint. The reduced dimensionality of the integral and the similarity of initial conditions in each trajectory pair also results in significant phase cancellation and an acceleration in convergence,
while the initial non-zero displacements in the quantum subsystem contribute phase
information essential to describing quantum coherence effects accurately.

Like MQC-IVR, AMQC-IVR is an approximate representation of the quantum mechanical time correlation function 
that offers mode-specific quantization in a dynamically uniform framework. 
While both methods offer a significant improvement over DHK-IVR's sign problem, AMQC-IVR does not require 
any tuning parameters, making it a more efficient and theoretically satisfying implementation of the mixed SC limit. 
Furthermore, as discussed in the following section, the AMQC-IVR prefactor is a good starting point for a variety of 
additional approximations that can improve computational efficiency even further.

\subsection{Separable Prefactor Approximation}\label{sec:seppref}
In the simplest case where the quantum and classical subsystems are not coupled, AMQC-IVR can be simplified significantly.
First, note that the phase terms and prefactors in the MQC-IVR and DHK-IVR integrands can be separated into the product 
of a quantum component and a classical component,
\begin{align}
G_t\parenth{\znot,\znotp;\bc}&e^{i\brak{S_t\parenth{\znot}-S_t\parenth{\znotp}}}=\nonumber\\
& G_t\parenth{\bz_Q^{},\bz_Q^\prime;\bc_Q}e^{i\brak{S_t\parenth{\bz_Q^{}}-S_t\parenth{\bz_Q^\prime}}}\nonumber\\
\times& G_t\parenth{\bz_C^{},\bz_C^\prime;\bc_C}e^{i\brak{S_t\parenth{\bz_C^{}}-S_t\parenth{\bz_C^\prime}}},\label{eq:sp1}\\
C_t^{}\parenth{\znot}C_t^*\parenth{\znotp}&e^{i\brak{S_t\parenth{\znot}-S_t\parenth{\znotp}}}=\nonumber\\
& C_t^{}\parenth{\bz_Q^{}}C_t^{*}\parenth{\bz_Q^\prime}e^{i\brak{S_t\parenth{\bz_Q^{}}-S_t\parenth{\bz_Q^\prime}}}\nonumber\\
\times& C_t^{}\parenth{\bz_C^{}}C_t^{*}\parenth{\bz_C^\prime}e^{i\brak{S_t\parenth{\bz_C^{}}-S_t\parenth{\bz_C^\prime}}}.
\label{eq:sp2}
\end{align}
Matrices $\bc_Q$ and $\bc_C$ represent the quantum and classical blocks of matrix $\bc$.
Given the limiting behavior of MQC-IVR as described in Sec.~\ref{sec:level2}, and given the separability of \eqn{eq:sp1}, it is straightforward to show that the AMQC-IVR expression for non-interacting quantum-classical subsystems is given by
\begin{align}
C_{AB}(t)=&\frac{1}{\parenth{2\pi}^{N+F}}\int d\znot\int d\bz_{Q}^\prime A_{\znot\bar{\bz}_0^{}}B_{\ztp\zt}\nonumber\\
\times &e^{i\brak{S_t\parenth{\bz_Q^{}}-S_t\parenth{\bz_Q^\prime}}}C_t^{}\parenth{\bz_Q^{}}C_t^*\parenth{\bz_Q^\prime}.
\label{eq:sp3}
\end{align}
Note that the phase terms and prefactors in \eqn{eq:sp3} only depend on the quantum subsystem. A comparison of \eqn{eq:sp3} and \eqn{eq:ML4} with the use of \eqn{eq:sp2} shows that, when quantum and classical subsystems do not interact, the AMQC-IVR prefactor reduces to unity,\footnote[3]{Note that, since the unprimed and primed trajectories of the classical subsystem are identical (when uncoupled from the quantum subsystem), the product of phase terms and prefactors associated with the classical subsystem is also equal to unity: $C_t^{}\parenth{\bz_C^{}}C_t^{*}\parenth{\bz_C^{}}e^{i\brak{S_t\parenth{\bz_C^{}}-S_t\parenth{\bz_C^{}}}}=1$.}
\begin{align}
\Lambda_t\parenth{\znot,\bz_Q^\prime}=1.
\label{eq:sp4}
\end{align}
We now use \eqn{eq:sp3} and \eqn{eq:sp4} to motivate an efficient approximation to AMQC-IVR when quantum and classical subsystems are \textit{weakly} coupled,
\begin{align}
C_{AB}(t)=&\frac{1}{\parenth{2\pi}^{N+F}}\int d\znot\int d\bz_Q^\prime A_{\znot\bar{\bz}_0^{}}B_{\zt\ztp}\nonumber\\
\times &e^{i\brak{S_t\parenth{\znot}-S_t\parenth{\bar{\bz}_0^{}}}}C_t^{}\parenth{\bz_Q^{}}C_t^*\parenth{\bz_Q^\prime}.
\label{eq:sp5}
\end{align}
We refer to \eqn{eq:sp5} as the SP approximation, and it is increasingly valid as the coupling between quantum and classical subsystems approaches zero. The advantage of the SP approximation is that it significantly reduces computational expense, particularly when the classical subsystem is larger than the quantum subsystem, i.e. when $N>>F$. This is because the SP approximation contains only two $F\times F$ Herman-Kluk prefactors, rather than two $N\times N$ Herman-Kluk prefactors and the additional $4N\times4N$ prefactor $\Lambda_t\parenth{\znot,\bz_Q^\prime}$ in \eqn{eq:ML4}.

There are several other approximations that could be made in addition to the SP approximation. For example, one could choose to evolve only a subset of the monodromy matrix elements, those associated with the quantum subsystem, albeit approximately, and reduce computation time even further. Furthermore, since the SP approximation contains only Herman-Kluk prefactors, which have been extensively studied, it is amenable to a variety of other representations and approximations\cite{gel00a,dil16a} that may reduce computational expense even further. In this study, we numerically
explore the applicability of the SP approximation, reserving other possibilities for future work.

\section{Model Systems}\label{sec:modsys}

Model 1 is a 1D anharmonic oscillator coupled to a heavy
harmonic `bath' mode. The Hamiltonian is given by
\begin{align}
\opH=&\frac{\hat{p}_1^2}{2m_1}+\frac{\hat{p}_2^2}{2m_2}+\onehalf m_1^{}\omega_1^2\hat{q}_1^2-0.1 \hat{q}_1^3+0.1\hat{q}_1^4\nonumber\\
&+\onehalf m_2^{}\omega_2^2\hat{q}_2^2+k\hat{q}_1\hat{q}_2,
\label{eq:MS1}
\end{align}
where $m_i$ and $\omega_i$ are the mass and frequency of the $i^{th}$ mode and $k$ is the bilinear coupling parameter. The initial state of the system is a product of coherent states, the position-space wavefunction of which is 
\begin{align}
\Psi\parenth{\bx,0}=&\braket{\bx|\bz_\mathbf{i}}\\
=&\mathcal{N}\prod_{j=1}^Ne^{-\frac{\gamma_{j}}{2}\parenth{x_j-q_{i_j}}^2+ip_{i_j}\parenth{x_j-q_{i_j}}},
\label{eq:MS2}
\end{align}
with $\mathcal{N}$ for normalization.

Model 2 contains the anharmonic mode of model 1 but now coupled to a bath of $N-1$ harmonic oscillators. The Hamiltonian is given by
\begin{align}
\opH=&\frac{\hat p_1^2}{2m_1}+\onehalf m_1\omega_1^2\hat q_1^2-0.1\hat q_1^3+0.1\hat q_1^4\nonumber\\
&+\sum_{j=2}^{N}\brak{\frac{\hat{p}_j^2}{2m_j}+\onehalf m_j\omega_j^2\parenth{\hat q_j-\frac{c_j\hat q_1}{m_j\omega_j^2}}^2}.
\label{eq:MS3}
\end{align}
The initial state of the full system is a product of $N$ coherent states, as in \eqn{eq:MS2}.

Model 3 is the widely used 1D symmetric double-well potential coupled to a thermal bath of $N-1$ harmonic oscillators. The Hamiltonian is
\begin{align}
\opH=&\frac{\hat p_1^2}{2m_1}+V\parenth{\hat{q_1}}\nonumber\\
&+\sum_{j=2}^{N}\brak{\frac{\hat{p}_j^2}{2m_j}+\onehalf m_j\omega_j^2\parenth{\hat q_j-\frac{c_j\hat q_1}{m_j\omega_j^2}}^2},
\label{eq:MS5}
\end{align}
with
\begin{align}
V(\hat q_1)=-\onehalf m_1\omega_b^2\hat{q}_1^2+\frac{m_1^2\omega_b^4}{16V_0^\ddagger}\hat{q}_1^4.
\label{eq:MS6}
\end{align}
In models 2 and 3, an Ohmic spectral density with an exponential cutoff is used for the bath,
\begin{align}
J(\omega)=\eta\omega e^{-\omega/\omega_c}.
\label{eq:MS7}
\end{align}
In model 2 we take $\omega_c=\omega_1$ and in model 3 we take $\omega_c=\omega_b$.

\section{Simulation Details}\label{sec:simdet}
With model 1 we take $\opB=\hat{x}_1$ to compute the average position of the anharmonic mode $\langle \hat x_1\rangle_t$ as a function of time.
We take $m_1=1.0$, $m_2=25.0$, $\omega_1=\sqrt{2}$, $\omega_2=1/3$, and 
we vary the bilinear coupling $k$ between $0.5$ and $2.0$, all in atomic units. 
Operator $\opA$ is the projection operator corresponding 
to the initial state,
\begin{align}
\opA=\ket{\bz_\mathbf{i}}\bra{\bz_\mathbf{i}}.
\label{eq:simd1}
\end{align}
The initial coherent states are centered at $q_{i_1}=q_{i_2}=1.0$ and $p_{i_1}=p_{i_2}=0.0$ with width parameters $\gamma_j=m_j\omega_j$ $\parenth{j=1,2}$. The initial conditions of the $j^{th}$ quantum dof are sampled from
\begin{align}
\rho\parenth{p_{0_j}^{},q_{0_j}^{},p_{0_j}^\prime,q_{0_j}^\prime}=\mathcal{N}&e^{-\frac{\gamma_j}{4}\parenth{q_{0_j}^{}-q_{i_j}}^2-\frac{1}{4\gamma_j}\parenth{p_{0_j}^{}-p_{i_j}}^2}\nonumber\\
\times&e^{-\frac{\gamma_j}{4}\parenth{q_{0_j}^\prime-q_{i_j}}^2-\frac{1}{4\gamma_j}\parenth{p_{0_j}^\prime-p_{i_j}}^2},\label{eq:samp1}
\end{align}
and the initial conditions of the $j^{th}$ classical dof are sampled from
\begin{align}
\rho\parenth{p_{0_j},q_{0_j}}=\mathcal{N}&e^{-\frac{\gamma_j}{2}\parenth{q_{0_j}^{}-q_{i_j}}^2}e^{-\frac{\gamma_j}{2}\parenth{p_{0_j}^{}-p_{i_j}}^2},\label{eq:samp2}
\end{align}
with $\mathcal{N}$ for normalization. The coherent state matrix element of $\hat x_1$ is
\begin{align}
\braket{\ztp|\hat x_1|\zt}=&\onehalf\brak{\parenth{q_{t_1}^\prime+q_{t_1}}-i\parenth{p_{t_1}^\prime-p_{t_1}}/\gamma_1}\nonumber\\
\times&\braket{\ztp|\zt}.
\end{align}
We found that approximately $10^5$ trajectory pairs are needed to converge AMQC-IVR calculations with model 1, whereas DHK-IVR required upwards of $10^8$. We also use a time step of $0.05$.

With model 2 we compute $\langle \hat x_1\rangle_t$ in the presence of $12$ bath modes as a function of time, and with different values of the coupling $\eta$ between the system and the bath. The initial coherent states are centered at $q_{i_1}=1.0$, $p_{i_1}=0.0$, and $q_{i_j}=p_{i_j}=0.0$ $\forall$ $j\in\brak{2,N=13}$. And the coherent state width parameter of the $j^{th}$ dof is given by $\gamma_j=m_j\omega_j$. The initial conditions of the quantum dof are sampled from \eqn{eq:samp1}, and the initial conditions of the classical dofs are sampled from \eqn{eq:samp2}. We found that AMQC-IVR requires approximately $10^6$ trajectory pairs for convergence. We also use a time step of $0.025$.
%

With model 3 we compute the thermal transmission coefficient $\kappa\parenth{T}$ of the symmetric double-well by means of a flux-side correlation function,\cite{mil93a}
\begin{align}
\kappa\parenth{T}=&\frac{k(T)}{k_{\text{cl}}^{\text{TST}}(T)}\\
=&\frac{1}{k_{\text{cl}}^{\text{TST}}(T)Q_r(T)}\lim_{t\rightarrow\infty}C_{fs}(t),
\end{align}
where $k_{\text{cl}}^{\text{TST}}(T)$ is the classical transition state theory result, $Q_r(T)$ is the partition function in the reactant well, and $C_{fs}(t)$ is the flux-side correlation function characterized by the following operators,
\begin{align}
\opA=&\,e^{-\beta\opH/2}\hat{F}e^{-\beta\opH/2},\label{eq:Aflux}\\
\opB=&\,\hat h.\label{eq:Bheav}
\end{align}
\eqn{eq:Aflux} contains the flux operator $\hat{F}=i\brak{\opH,\hat h}$ and constant $\beta=1/kT$ with temperature $T=300K$ and Boltzmann constant $k$. In \eqn{eq:Bheav} $\hat h$ is the unit step function specifying the dividing surface. Numerical parameters for the Hamiltonian of model 3 can be found in Ref.~[22]. In order to evaluate the coherent state matrix element of $\opA$ in \eqn{eq:Aflux} we make a normal-mode approximation at the transition state\cite{wan00a} so that the Hamiltonian is approximately separable:
\begin{align}
\opH\approx&\opH_1+\sum_{j=2}^{N=13}\opH_j,\\
\opH_1=&\frac{\hat{p}_1^2}{2m}-\onehalf m\lambda^{\ddagger2}\hat x_1^2+V_0^\ddagger,\\
\opH_j=&\frac{\hat p_j^2}{2m_j}+\onehalf m_j\lambda_j^2\hat{x}_j^2.
\end{align}
Frequencies $\lambda^{\ddagger}$ and $\lambda_j$ are the imaginary and real normal-mode frequencies at the transition state, respectively. Under this approximation, the coherent state matrix element of $\opA$ in \eqn{eq:Aflux} is given by
\begin{align}
A_{\znot\znotp}=&\mathcal{F}_{z_{Q}^{}z_{Q}^\prime}\prod_j\parenth{\mathcal{B}_{z_C^{}z_C^\prime}}_j,\label{eq:sampler1}\\
\mathcal{F}_{z_{Q}^{}z_{Q}^\prime}=&\frac{\gamma_1}{8m_1\sqrt{\pi}\cos^2{u^\ddagger}}\nonumber\\
\times &\brak{\parenth{p_{0_1}^\prime+p_{0_1}^{}}/\sqrt{\gamma_1}-i\sqrt{\gamma_1}\parenth{q_{0_1}^\prime-q_{0_1}^{}}}\nonumber\\
\times & e^{-\frac{\gamma_1}{4}\parenth{q_{0_1}^{\prime2}+q_{0_1}^2}}e^{-\frac{1}{4\gamma_1}\parenth{p_{0_1}^{\prime2}+p_{0_1}^2}}\nonumber\\
\times & e^{\frac{i}{2}\parenth{p_{0_1}^{}+p_{0_1}^\prime}\parenth{q_{0_1}^{}-q_{0_1}^\prime}}e^{-\beta V_0^\ddagger},\\
\parenth{\mathcal{B}_{z_C^{}z_C^\prime}}_j=&e^{-\frac{\gamma_j}{4}\parenth{q_{0_j}^2+q_{0_j}^{\prime2}}}e^{-\frac{1}{4\gamma_j}\parenth{p_{0_j}^2+p_{0_j}^{\prime2}}}e^{\frac{i}{2}\parenth{p_{0_j}^{}q_{0_j}^{}-p_{0_j}^\prime q_{0_j}^\prime}}\nonumber\\
\times & e^{\onehalf e^{-2u_j}\brak{\gamma_j q_{0_j}^{} q_{0_j}^\prime+p_{0_j}^{}p_{0_j}^\prime/\gamma_j+i\parenth{p_{0_j}^\prime q_{0_j}^{}-p_{0_j}^{} q_{0_j}^\prime}}}e^{-u_j},\label{eq:bathB}
\end{align}
with $u_1=\beta\left|\lambda^\ddagger\right|/2$, $u_j=\beta\lambda_j/2$ $\forall$ $j$ $=$ $2\dots N$, and we choose $\gamma_1=m_1|\lambda^\ddagger|\cot u^\ddagger$ as well as $\gamma_j=m_j\lambda_j$ $\forall$ $j$ $=$ $2\dots N$. With AMQC-IVR, however, only the diagonal elements of \eqn{eq:bathB} are needed, i.e.
\begin{align}
&\braket{\znot|\opA|\bar{\bz}_0^{}}=\mathcal{F}_{z_{Q}^{}z_{Q}^\prime}\prod_j\parenth{\mathcal{B}_{z_{C}^{}z_{C}^{}}}_j,\label{eq:sampler2}\\
&\parenth{\mathcal{B}_{z_{C}^{}z_{C}^{}}}_j=e^{-\frac{\gamma_j}{2}\parenth{1-e^{-2u_j}}q_{0_j}^2}e^{-\frac{1}{2\gamma_j}\parenth{1-e^{-2u_j}}p_{0_j}^2}e^{-u_j}.
\end{align}
Note that the products on the right-hand sides of \eqn{eq:sampler1} and \eqn{eq:sampler2} are over all components of the classical subsystem.
Initial conditions for the quantum and classical dofs are sampled from the following distributions,
\begin{align}
\rho\parenth{q_{0_1}^{},p_{0_1}^{},q_{0_1}^\prime,p_{0_1}^\prime}=\mathcal{N}&e^{-\frac{\gamma_1}{4}\parenth{q_{0_1}^{\prime2}+q_{0_1}^2}}\nonumber\\
\times &e^{-\frac{1}{4\gamma_1}\parenth{p_{0_1}^{\prime2}+p_{0_1}^2}},\\
\rho\parenth{q_{0_j},p_{0_j}}=\mathcal{N}&\parenth{\mathcal{B}_{z_{C}^{}z_{C}^{}}}_je^{u_j},
\end{align}
respectively. Constant $\mathcal{N}$ is for normalization. The coherent state matrix element of $\hat h$ at time $t$ is given by
\begin{align}
h_{\ztp\zt}=&\onehalf\brak{\text{Erf}\parenth{\onehalf\alpha_t}+1}\braket{\ztp|\zt},\\
\alpha_t=&\sqrt{\gamma_1}\parenth{q_{t_1}^{}+q_{t_1}^\prime}-i\parenth{p_{t_1}^\prime-p_{t_1}^{}}/\sqrt{\gamma_1}.
\end{align}
$\text{Erf}\parenth{x}$ is the error function of $x$. We found that approximately $10^6$ trajectory pairs were required for convergence. We also use time step of $10.0$. The SC Corr-Code Package, an open-source program developed in-house, was used to run all the simulations in this study.\cite{corrcode}

\section{Results and Discussion}\label{sec:res}
\begin{figure}
\includegraphics[scale=0.6]{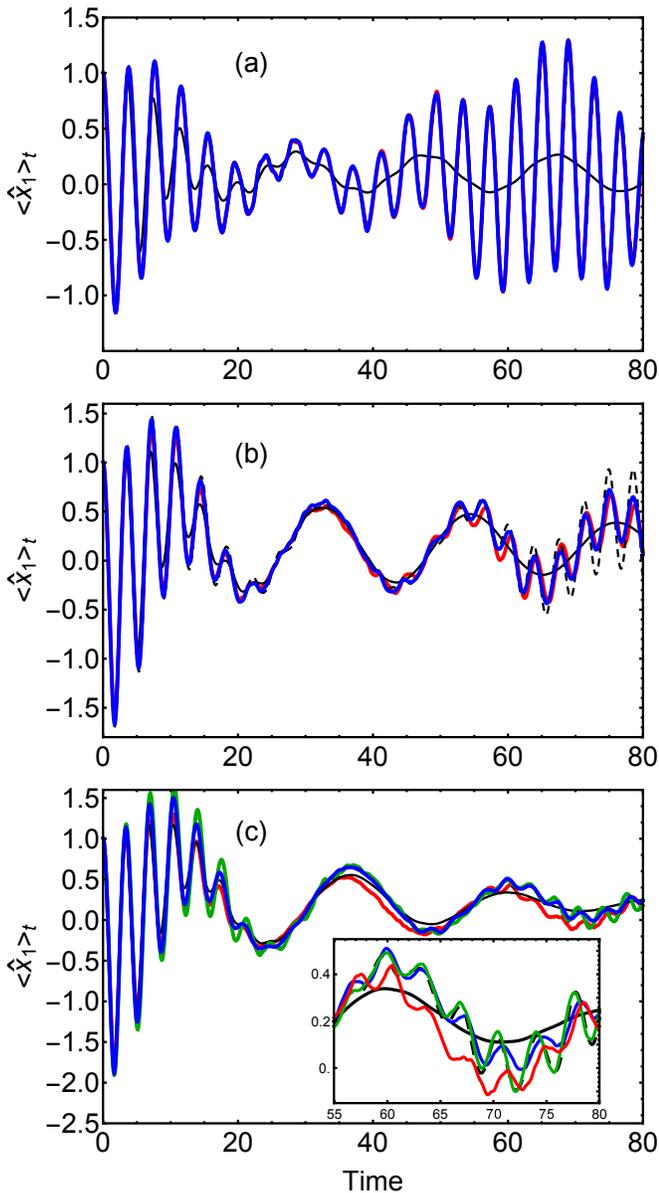}
\caption{The average position of the anharmonic mode in model 1 as a function of time, as computed with exact quantum (black, dashed), Husimi-IVR (black), as well as AMQC-IVR with one quantized mode (blue), AMQC-IVR with one quantized mode under the SP approximation (red), and, in (c), AMQC-IVR with two quantized modes (green). Each panel corresponds to a different coupling strength used during the simulation: 
(a) $k=0.5$, (b) $k=1.5$, and (c) $k=2.0$. The inset in (c) amplifies the correlation function from $t=55$ to $t=80$.}
\label{fig:fig1}
\end{figure}

The position expectation value of the anharmonic mode in model 1 is plotted as a function of time in \figr{fig:fig1} 
with three different coupling strengths. The rapid oscillations in the exact quantum results (dashed) at long times 
are a result of nuclear coherence, a feature that is clearly absent in the classical limit Husimi-IVR results (black). 
The AMQC-IVR results (blue), in which the anharmonic mode is treated in the quantum limit and the harmonic mode is treated 
in the classical limit, is consistently accurate (see \tabl{tab:err1}) with only slight damping in the coherences
at long times when the quantum and classical modes are strongly coupled. 
\begin{table}[h]
\begin{tabular}{| c || c | c | c |}\hline
$k$	&	$0.5$	&	$1.5$ 	&	$2.0$	\\\hline
Avg. \% Error	&	$1.07$	&	$0.96$	&	$0.91 (0.31)$\\\hline
\end{tabular}
\caption{The time-averaged (relative) \% error of each AMQC-IVR result with model 1. The result in parentheses was obtained after quantizing both the anharmonic and harmonic dofs.}
\label{tab:err1}
\end{table}

Also plotted in \figr{fig:fig1} are the AMQC-IVR results obtained with the SP approximation (red) of \eqn{eq:sp5}. 
In \figr{fig:fig1}(a) and \figr{fig:fig1}(b) these results are nearly identical to the AMQC-IVR results without 
approximations to the prefactor. There is some additional loss of long-time coherences in the SP approximation in \figr{fig:fig1}(c), 
and the reported error increases (see \tabl{tab:err2}).  We note that the SP results with this low-dimensional model serve only as a proof of principle, offering
limited reduction in computational cost.
\begin{table}[h]
\begin{tabular}{| c || c | c | c | c |}\hline
$k$	&	$0.5$	&	$1.5$ 	&	$2.0$	\\\hline
Avg. \% Error	&	$1.06$	&	$1.53$	&	$1.85$\\\hline
\end{tabular}
\caption{The time-averaged (relative) \% error of the SP approximation with model 1.}
\label{tab:err2}
\end{table}

\figr{fig:fig1}(c) also plots $\langle \hat{x}_1\rangle_t$ as computed with AMQC-IVR 
when both modes are treated in the quantum limit. Since all dofs are quantized here, 
this is equivalent to using DHK-IVR of \eqn{eq:DHK}, but the AMQC-IVR prefactors were 
used during computation. The result is nearly identical to the exact quantum result 
at all times, showing that, by increasing the size of the quantum subsystem, 
AMQC-IVR results can be systematically improved toward the DHK-IVR limit.

\begin{figure}
\includegraphics[scale=0.6]{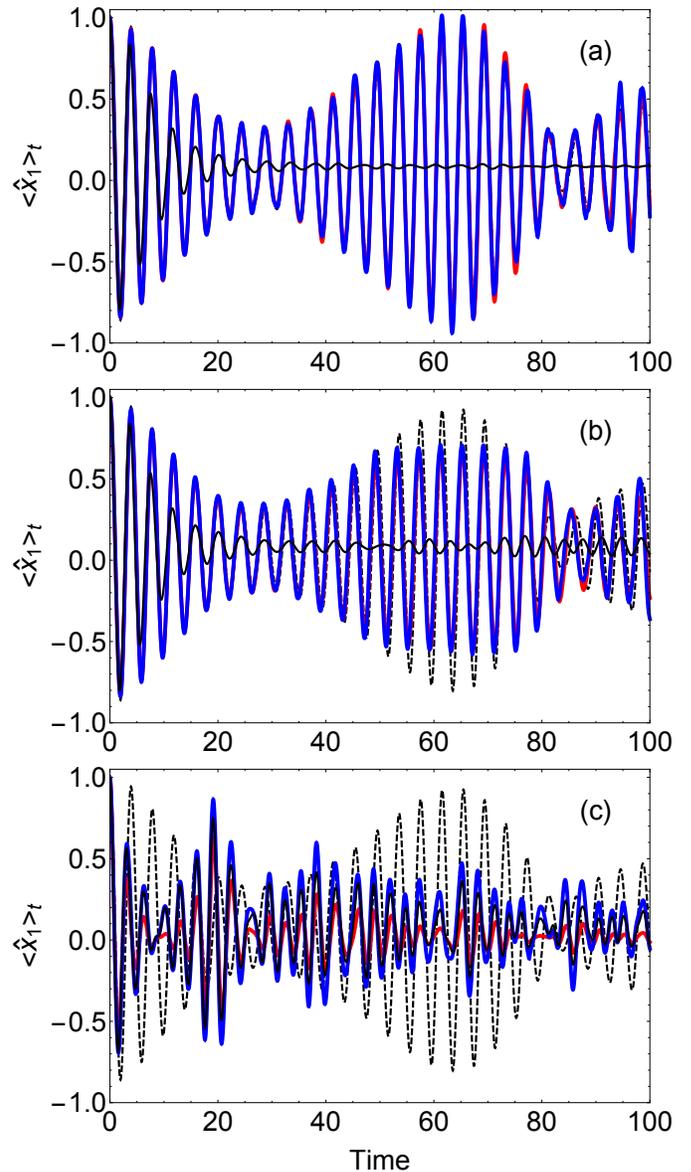}
\caption{The average position of the anharmonic mode in model 2 as computed with AMQC-IVR (blue), the SP approximation (red), and Husimi-IVR (black, solid) with a coupling strength of (a) $\eta/m\omega_1=10^{-4}$, (b) $\eta/m\omega_1=10^{-2}$, and (c) $\eta/m\omega_1=1.0$. The black dashed curve is the exact quantum result of the 1D anharmonic oscillator in the absence of coupling to the bath.}
\label{fig:fig2}
\end{figure}

The average position of the anharmonic mode of model 2 is plotted in \figr{fig:fig2} as a function of time. Each panel uses a different reduced coupling strength $\eta/m\omega_1$ between the anharmonic dof and the harmonic bath. And each AMQC-IVR result was obtained by quantizing the anharmonic mode and treating the harmonic bath in the classical limit.
In \figr{fig:fig2}(a), where the coupling between the system and bath is very weak, 
AMQC-IVR (blue) is nearly identical to the exact result of the uncoupled anharmonic 
oscillator (dashed), but with a slight overestimation of the amplitudes at later times. 
The SP approximation (red) in \figr{fig:fig2}(a) is nearly identical the 
exact uncoupled result as well, and computational time was reduced by a factor of four.
The classical limit result obtained with Husimi-IVR (black) in \figr{fig:fig2}(a) appears to fail at all but very short times. A comparison of these results suggest that AMQC-IVR and the SP approximation are accurately capturing nuclear coherence effects in the position correlation function. This is encouraging since a full DHK-IVR treatment of such a highly multidimensional system is not possible without a remedy to the sign problem.

In \figr{fig:fig2}(b), where the coupling between the system and bath is $100$ times stronger than in \figr{fig:fig2}(a), the AMQC-IVR and SP results again resemble the oscillatory structure of the uncoupled result, albeit with damped amplitudes around $t=65$ due to the stronger influence from bath dofs. 
Once again, the Husimi-IVR result in \figr{fig:fig2}(b) does not show the oscillatory behavior 
of the AMQC-IVR with and without the SP approximation.

\figr{fig:fig2}(c), where the coupling is $10^4$ times stronger than in \figr{fig:fig2}(a), the AMQC-IVR and Husimi-IVR results are very similar, though the amplitudes in the AMQC-IVR result are slightly larger from about $t=20$ and beyond. Since the coupling between the system and bath is strong here, one would expect that the coherence effects associated with the motion of the system would be significantly damped due the strong influence from the bath. It is therefore reasonable to see that AMQC-IVR and Husimi-IVR give similar results in the strong coupling limit. The oscillatory structure of the SP approximation in \figr{fig:fig2}(c) resembles that of the AMQC-IVR and Husimi-IVR results, but with some moderate damping of the amplitudes, particularly at longer times. However, since the SP approximation is most valid when the coupling between the system and bath is weak, it is intuitive that the SP approximation is less reliable than AMQC-IVR in \figr{fig:fig2}(c).

The thermal transmission coefficient of the symmetric double well of model 3 is plotted in \figr{fig:fig3} as a function of the reduced coupling strength $\eta/m\omega_b$ (see Appendix C for the tabulated data). AMQC-IVR results (red) were obtained by quantizing the double-well and treating each bath dof in the classical limit.
\begin{figure}
\includegraphics[scale=0.6]{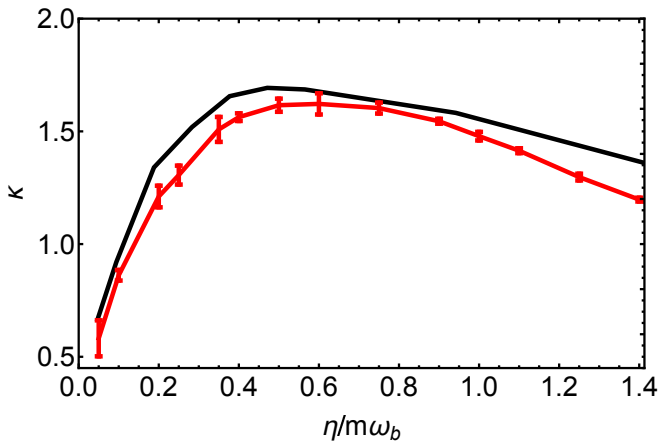}
\caption{The thermal transmission coefficient of model 3 at $T=300K$ as a function of the reduced coupling strength. Exact path integral results\protect\cite{top94a} were obtained with permission from Ref.~[74].}
\label{fig:fig3}
\end{figure}
AMQC-IVR clearly agrees well with the exact quantum results (black) in the weak coupling limit, 
and captures the turnover region around $\eta/m\omega_b=0.5$. In the stronger coupling limit, 
however, AMQC-IVR begins to increasingly underestimate the transmission coefficient, though the 
trend is qualitatively correct. It is reasonable to expect AMQC-IVR to fail in the strong 
coupling limit, since much of the phase information from bath dofs is removed. 

\begin{figure}
\includegraphics[scale=0.6]{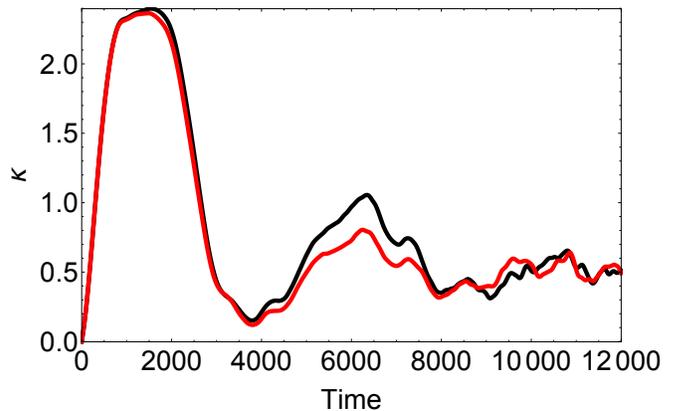}
\caption{The thermal transmission coefficient of model 3 at $T=300K$ as a function of time, as computed with AMQC-IVR (black) and the SP approximation (red), for a fixed weak coupling strength of $\eta/m\omega_b=0.05$.}
\label{fig:fig4}
\end{figure}

\figr{fig:fig4} plots the thermal transmission coefficient of model 3 in the 
weak coupling regime, $\eta/m\omega_b=0.05$,
as a function of time with AMQC-IVR (black) and the SP approximation (red). 
Both results show an oscillatory structure in the transmission coefficient, 
corresponding to the transfer of population between the two wells. 
The amplitude of the second peak ($t=6000$) in the SP result is slightly damped 
relative to the AMQC-IVR result, but, appealingly, both results clearly converge 
to the same long-time limit, where the rate is determined. Furthermore, computation 
with the SP approximation was $4$ times faster than without, and when the number of 
bath modes was doubled to $24$ (which does not affect the rate), computation was $8$ times 
faster than without. Note, however, that while the $\%$ error between AMQC-IVR and the SP 
approximation is small when $\eta/m\omega_b=0.05$, the SP approximation increasingly 
underestimates the rate for larger coupling strengths (see Table~\ref{tab:err3}).
\begin{table}[h]
\begin{tabular}{| c | c | c | c |}\hline
$\eta/m\omega_b$	&	$0.05$	&	$0.1$	&	$0.5$	\\\hline\hline
$\% \text{Error}$		&	$3.0$	&	$38$		&	$65$		\\\hline
\end{tabular}
\caption{The percent error between the AMQC-IVR and SP results for the thermal transmission coefficient at three different coupling strengths.}
\label{tab:err3}
\end{table}
Overall, our analysis of model 3 shows that AMQC-IVR and the SP approximation can be very reliable in computing chemical reaction rates in the condensed phase, particularly 
when the coupling between quantum and classical subsystems is weak.

\section{Conclusions}\label{sec:conc}
An analytical parameter-free mixed quantum-classical limit of DHK-IVR has been 
derived such that some modes of the system can be treated in the quantum 
and others in the classical limit of SC-IVR theory. AMQC-IVR shares the 
advantages of MQC-IVR as a methodology that both mitigates the SC-IVR sign problem 
and offers mode-specific quantization in a dynamically uniform framework. 
But AMQC-IVR offers a reduction in dimensionality of the phase space integral
and it removes the task of having to choose an optimal set of tuning parameters. 
We have tested AMQC-IVR on three system-bath models and showcased its numerical 
accuracy across coupling regimes. We have also shown that the SP approximation to AMQC-IVR 
can be very accurate when the coupling between quantum and classical dofs is weak, 
and very efficient when the quantum subsystem is smaller than the classical subsystem.

\section*{Acknowledgements}
The authors thank Prof. Makri for providing the exact quantum results for the 
transmission coefficient used in \figr{fig:fig3}. This work was funded by an NSF CAREER Grant No. CHE 1555205
and the Research Corporation for Science Advancement through a Cottrell Scholar Award.
\appendix
\section{MQC-IVR Prefactor}
\label{ap:dfp}
The MQC-IVR\cite{chu17a} prefactor is given by
\begin{align}
&D_t\parenth{\znot,\znotp;\cpp,\cq}=\det\big[\onehalf \gamt^{-1}\Gmat\big]\Ronehalf\det\big[\onehalf\Amat\parenth{\Gmat^{-1}+\Imat}\Bmat\nonumber\\
&+\Cmat\parenth{\onehalf\gamz^{-1}+\cpp}\Gmat^{-1}\Bmat+\onehalf\Cmat\parenth{\Gmat^{-1}+\Imat}\Dmat\nonumber\\
&+\Amat\parenth{\onehalf\gamz+\cq}\Gmat^{-1}\Dmat\big]\Ronehalf,
\end{align}
with
\begin{align}
\Amat=&\mppf-i\gamt\mqpf\\
\Bmat=&\mppbb\gamt+i\mpqbb\\
\Cmat=&\gamt\mqqf+i\mpqf\\
\Dmat=&\mqqbb-i\mqpbb\gamt\\
\Gmat=&\parenth{\cq+\gamz}\cpp+\cq\parenth{\gamz^{-1}+\cpp}.
\end{align}

\section{Derivation of AMQC-IVR}
\label{ap:mlp}
Here we derive AMQC-IVR in detail. According to a previous study\cite{chu17a} one can rewrite \eqn{eq:ML3} as,
\begin{widetext}
\begin{align}
\lim_{c_{\text{quantum}}\rightarrow0\atop c_{\text{classical}}\rightarrow\infty}G_t\parenth{\znot,\znotp;\bc}=&\parenth{2\pi}^{N-F}\bar D_t\parenth{\znot,\znotp}\prod_{j=F+1\atop j=N+F+1}^{2N\atop N}\delta\parenth{\Delta_{0_j}},\\
\bar D_t\parenth{\znot,\znotp}=&\lim_{c_{\text{quantum}}\rightarrow 0\atop c_{\text{classical}}\rightarrow\infty}\prod_{j=F+1\atop j=N+F+1}^{2N\atop N}\brak{\frac{1}{c_{jj}}}\Ronehalf\det\brak{\Kmat\T+i\tilde{\bc}\Jmat}\Ronehalf.
\label{eq:appB2}
\end{align}
\end{widetext}
$\Kmat\T$ is a complex $4N\times4N$ matrix,
\begin{align}
\Kmat\T=&
\begin{pmatrix}
\Xmat^{}	&	\Xmat^{*}	\\
\Ymat\Mmat^\prime	&	\Ymat^*\Mmat
\end{pmatrix},
\label{eq:appB3}
\end{align}
with constant $2N\times 2N$ matrices
\begin{align}
\Xmat=&
\begin{pmatrix}
\frac{i}{2}\gamz	&	-\onehalf\Imat	\\
\onehalf\Imat	&	\frac{i}{2}\gamz^{-1}
\end{pmatrix},\\
\Ymat=&
\begin{pmatrix}
\frac{i}{2}\gamt	&	\onehalf\Imat	\\
-\onehalf\Imat	&	\frac{i}{2}\gamt^{-1}
\end{pmatrix}.
\label{eq:appB4}
\end{align}
$\Mmat$ and $\Mmat^\prime$ are the full $2N\times2N$ monodromy matrices for trajectories beginning at $\znot$ and $\znotp$, respectively.
We also have the $4N\times4N$ diagonal matrix of tuning parameters,
\begin{align}
\tilde{\bc}=&
\begin{pmatrix}
\bc	&	\nullmat	\\
\nullmat	&	\nullmat	
\end{pmatrix},
\label{eq:appB5}
\end{align}
with $\bc$ defined in \eqn{eq:cmatrix1}. We also have the
$4N\times4N$ matrix $\Jmat$,
\begin{align}
\Jmat=
\begin{pmatrix}
\Imat	&	-\Imat	\\
\Mmat^\prime	&	-\Mmat
\end{pmatrix}.
\label{eq:appB6}
\end{align}
The procedure from here is to expand the determinant on the right-hand side of \eqn{eq:appB2} and evaluate the limit term-by-term. Using the definition of the determinant of a general $4N\times 4N$ matrix
$\Ximat$,
\begin{align}
\det\brak{\Ximat}=\sum_{i_1,\dots,i_{4N}}^{4N}\epsilon_{i_1\dots i_{4N}}\prod_{j=1}^{4N}\Xi_{ji_j},
\label{eq:appB7}
\end{align}
where $\epsilon$ is the Levi-Civita symbol, the expansion of the determinant in \eqn{eq:appB2} 
gives
\begin{widetext}
\begin{align}
\bar{D}_t\parenth{\znot,\znotp}=&\lim_{c_{\text{quantum}}\rightarrow 0\atop c_{\text{classical}}\rightarrow\infty}\bigg\{\prod_{j=F+1\atop j=N+F+1}^{2N\atop N}\brak{\frac{1}{c_{jj}}}\sum_{i_1,\dots,i_{4N}}^{4N}\epsilon_{i_1\dots i_{4N}}\prod_{j=1}^{4N}\brak{\Kmat\T+i\tilde\bc\Jmat}_{ji_j}\bigg\}\Ronehalf.
\label{eq:appB8}
\end{align}
\end{widetext}
Given that $\tilde{\bc}$ is diagonal, and given that the lower blocks of $\tilde{\bc}$ contain only zeros, we have
\begin{widetext}
\begin{align}
&\bar{D}_t\parenth{\znot,\znotp}=\lim_{c_{\text{quantum}}\rightarrow 0\atop c_{\text{classical}}\rightarrow\infty}\bigg\{\prod_{j=F+1\atop j=N+F+1}^{2N\atop N}\brak{\frac{1}{c_{jj}}}\sum_{i_1,\dots,i_{4N}}^{4N}\epsilon_{i_1\dots i_{4N}}\prod_{j=1}^{2N}\brak{K_{ji_j}\T+i{c}_{jj}J_{ji_j}}\prod_{j=2N+1}^{4N}\brak{K_{ji_j}\T}\bigg\}\Ronehalf.
\label{eq:appB10}
\end{align}
\end{widetext}
When the leading product on the right-hand side of \eqn{eq:appB10} is carried through the summation, the only terms that will survive the limit are those that are independent of the elements of $\bc$. We then have,
\begin{widetext}
\begin{align}
\bar{D}_t\parenth{\znot,\znotp}=&\bigg\{\sum_{i_1,\dots,i_{4N}}^{4N}\epsilon_{i_1\dots i_{4N}}\prod_{j=F+1\atop j=N+F+1}^{2N\atop N}\prod_{m\ne j}^{4N}\brak{iJ_{ji_j}}\brak{K_{mi_m}\T}\bigg\}\Ronehalf.
\label{eq:appB11}
\end{align}
\end{widetext}
Note that the right-most product on the right-hand side of \eqn{eq:appB11} is over all $m$ $\in\brak{1,F}$, all $m$ $\in\brak{N+1,N+F}$, and all $m$ $\in\brak{2N+1,4N}$.
The right-hand side \eqn{eq:appB11} can be viewed as the determinant of matrix $\Kmat\T$ plus an additional matrix $\mathbf{\Sigma}$,
\begin{align}
\bar D_t\parenth{\znot,\znotp}=\det\brak{\Kmat\T+\mathbf{\Sigma}}\Ronehalf.
\label{eq:appB11a}
\end{align}
The addition of matrix $\mathbf{\Sigma}$ to matrix $\Kmat\T$ effectively replaces certain rows of $\Kmat\T$ (specifically, rows associated with the classical subsystem) with
the corresponding rows of matrix $i\Jmat$. Therefore,
\begin{align}
\mathbf{\Sigma}=&
\begin{pmatrix}
\Ommat	&	\Ommat^*	\\
\nullmat	&	\nullmat
\end{pmatrix},\label{eq:appB12}
\\
\parenth{\Ommat}_{jk}=&
\begin{cases}
i\parenth{1-\onehalf\gamma_{jj}}	  &	F<j\leq N,\\
i\parenth{1-\onehalf\gamma_{jj}^{-1}}  &	N+F<j\leq 2N,\\
-\onehalf\delta_{j-N,k}				  &    	N+F< j \leq 2N, F< k\leq N,\\
\onehalf\delta_{j,k-N}				  &	F< j\leq N, N+F< k\leq 2N\\
0	&	\text{else}.
\end{cases}
\label{eq:appB13}
\end{align}
Since matrix $\Kmat\T$ is invertible, we can equivalently write \eqn{eq:appB11a} as
\begin{align}
\bar D_t\parenth{\znot,\znotp}=\det\brak{\Kmat\T}\Ronehalf\det\brak{\Imat+\mathbf{\Sigma}\parenth{\Kmat\T}^{-1}}\Ronehalf.
\label{eq:appB14}
\end{align}
And since the leading term on the right-hand side of \eqn{eq:appB14} is proportional to the product of HK prefactors,
\begin{align}
\det\brak{\Kmat\T}\Ronehalf=\parenth{-1}^\frac{N}{2}C_t\parenth{\znot}C_t^*\parenth{\znotp},
\label{eq:appB15}
\end{align}
the AMQC-IVR prefactor is given by,
\begin{align}
\Lammat_t\parenth{\znot,\znotp}=\parenth{-1}^\frac{N}{2}\det\brak{\Imat+\mathbf{\Sigma}\parenth{\Kmat\T}^{-1}}\Ronehalf.
\label{eq:appB16}
\end{align}
Note that, for the model systems studied here, however, \eqn{eq:appB11a} was used for all computations. The results of \eqn{eq:appB14}, \eqn{eq:appB15}, and \eqn{eq:appB16} are used here to motivate the SP approximation.

\section{Tabulated Thermal Transmission Coefficients}
\begin{table}[h!]
\begin{tabular}{| c || c |}\hline
$\eta/m\omega_b$	&	$\kappa$	\\\hline\hline
0.05		&	0.58(7)	\\\hline
0.1		&	0.86(2)	\\\hline
0.2		&	1.21(4)	\\\hline
0.25		&	1.31(4)	\\\hline
0.35		&	1.51(6)	\\\hline
0.4		&	1.56(2)	\\\hline
0.5		&	1.62(3)	\\\hline
0.6 		&	1.62(5)	\\\hline
0.75 		&	1.60(2)	\\\hline
0.9 		&	1.54(1)	\\\hline
1.0 		&	1.48(2)	\\\hline
1.1 		&	1.41(1)	\\\hline
1.25		&	1.30(2)	\\\hline
1.4		&	1.20(1)	\\\hline
1.5		&	1.12(1)	\\\hline
\end{tabular}
\caption{The AMQC-IVR results for the thermal transmission coefficient of model 3.}
\end{table}

\newpage

\bibliography{mybib}

\end{document}